# A SMART WIZARD SYSTEM SUITABLE FOR USE WITH INTERNET MOBILE DEVICES TO ADJUST PERSONAL INFORMATION PRIVACY SETTINGS

Nahier Aldhafferi, Charles Watson and A.S.M Sajeev

School of Science and Technology, University of New England, Australia
naldaffe@myune.edu.au

## ABSTRACT

*The privacy of personal information is an important issue affecting the confidence of internet users. The widespread adoption of online social networks and access to these platforms using mobile devices has encouraged developers to make the systems and interfaces acceptable to users who seek privacy. The aim of this study is to test a wizard that allows users to control the sharing of personal information with others. We also assess the concerns of users in terms of such sharing such as whether to hide personal data in current online social network accounts. Survey results showed the wizard worked very well and that females concealed more personal information than did males. In addition, most users who were concerned about misuse of personal information hid those items. The results can be used to upgrade current privacy systems or to design new systems that work on mobile internet devices. The system can also be used to save time when setting personal privacy settings and makes users more aware of items that will be shared with others.*

## 1. INTRODUCTION

Personal information privacy has become a big challenge facing developers of internet applications. It is difficult to guarantee 100% privacy of personal information especially on online social network sites. Personal information can be misused by anonymous individuals, friends, or others, and no single concern is responsible for protecting such data. Many features must work smoothly to increase the degree of privacy protection. For example, user awareness, new privacy applications, and the use of simple tools to control privacy settings, decrease risks. In this article, we develop a new wizard to help users control the privacy level of their personal information. We describe the implementation thereof and survey user acceptability.

## 2. LITERATURE REVIEW

### 2.1. Social network sites

In the early days online social network sites were used mostly among college students. Today, no such limitation exists and various groups in society use social networking to communicate or to read news updates. Further, advertising on such sites has become common. Competition between sites is vigorous; the sites strive to continually add new services to attract users. As those who use such sites are predominantly young [1], governments, organizations, and institutions have developed special laws to protect user data [2].

Although networking is popular worldwide, no website dominates the scene. For example, Facebook and Myspace are commonly used in the US and Australia but Cyworld is more popular in Korea and Orkut is preferred in Brazil and India. Additionally, each online social network website has an individual privacy system; the specific features vary. Some websites have complicated privacy settings whereas others do not [3].

## 2.2. Privacy and risk in social networking

Recently, it has become clear that the use of online social networks is not limited to adults; high proportions of children in some countries also have accounts. In [4] authors found that about 77% of European children 13–16 years of age have profiles in at least one social network. The survey spanned 25 countries and, in each country, most users were within this age bracket Also, 38% of children aged 9–12 years have social network accounts; this indicates that social network usage will become even more common.

The cited authors examined parental restrictions on use of networks by children. About 32% of children are not supervised and about 20% are supervised to some extent. About half of all parents did not restrict the activities of their children. The survey did not study whether restrictions were related to privacy concerns or whether other considerations were in play.

The survey of Ai Ho, Maiga and Aimeuer [5], which included 200 participants, revealed some problems with privacy issues. The most pressing issue was that sites did not clearly inform users of the risk that divulged personal information might be misused. Also, few tools were available to protect personal information. Finally, users could not control what others might publish about them.

- The cited authors classified personal information into five categories:
- Identification: Details identifying a user, such as a name, address, or telephone number.
- Demography: Any personal information on appearance or characteristics such as age, gender, weight, or political view.
- Activity: User activities such as writing of comments, addition of friends, or changing of current status.
- Social networking: Relationships between the user and others on the network, such as friends.
- Added content: Pictures, videos, and music.

The large number of social network users may stimulate an increase in the number of malicious attacks [6] thus affecting privacy in various ways. APIs (Application Programming Interfaces) may violate user privacy. Allowing such applications to run may allow third-party access to personal information; application developers can access user data. Thus, social network hosts must protect user data and supervise all APIs requesting access to such data [7].

Use of online social networking offers many benefits. Friends and jobs may be found, interests and information shared, and comments exchanged. However, the placement of ever more personal information on such sites can create privacy risks for some users; this is particularly the case if a user is not sophisticated [8]. In [9] authors emphasized that protection of user privacy is a responsibility of the service provider.

## 2.3. Privacy of personal information

Many challenges face those who try to ensure privacy; design of a user interface that can be used to control privacy settings may be helpful in this regard. In [10] authors discussed several points requiring consideration in this context:

- Collection of personal data must be limited.
- The purpose of data collection and use of such data must be defined.
- Data must not be used for any other purpose.
- Personal information must be protected and stored securely.
- The rights and duties of users and those who view data must be defined.
- Accountability principles must be developed to confirm that these principles are applied and to continuously monitor compliance.

The current privacy systems of online social networks are not trustworthy when millions are on the system [11]. Over 25% of children aged 9–16 years old set their profile pages to "public", allowing general viewing [4]. Thus, various laws have been promulgated to protect the personal information of children. In 1998, the US passed a federal Children's Online Privacy Protection law applicable to those 13 years of age or younger [30]. No website may collect personal data from children unless parents so permit. However, the commercial pressures are strong; children are receptive to specific advertisements and high-school details (for example) would be of value to advertisers and college recruiters [12]. These authors also explained that although some websites such as Facebook and Google+ seek to comply with the law by preventing all children under 13 years of age from registering, some game the system by giving false birthdates. Websites such as Facebook do apply privacy policies relevant to minors. For example, children can receive messages only from their own friends or from people who give contact details such as an email address or a phone number, but adults can receive messages from anyone. These restrictions also apply (inter alia) to posting of pictures and addition of friends [13]. Baden and others developed a technique termed Persona which can be used to hide personal information by combining attribute-based encryption with a traditional public access key [11].

In recent times, most users have (somewhat) restricted access to their online social network profiles. In a study conducted by Madden [14], about 58% of users (48% of males and 67% of females) allow their profiles to be seen only by friends. Males restrict access to a limited extent, but females are more concerned about privacy profiles. Also, about 67% of females have deleted some people from their friends lists compared to about 58% of males. However, the ability to control privacy settings varies among networks. About 48% of participants found it difficult to change settings; the ease of changing improved with higher educational level.
In [4] authors found that the website "Hyves" was rated highly by most users in terms of the availability and ease of use of privacy features. Privacy settings were easily changed and other users blocked.

## 2.4. Usability of internet mobile devices

The widespread use of mobile internet devices has encouraged companies and internet software developers to prioritize usability. Several challenges are apparent. Some are hardware-based and the others software issues. One important question is screen size, image fit differs between desktop computers and mobile phones, and, of course, phones do not have a full-sized keyboard or mouse. Various companies have developed internet-specific mobile browsers including Opera, Internet Explorer, and Safari [15]. Today, most mobile internet browsers support various programming languages including HTML and JavaScript, but they do not browse as effectively as do laptops or PCs because the screens and keyboards are smaller [16].

Developments in this field are not confined to commercial purposes. In the educational field, for example, in [17] authors designed a tool allowing quizzes to be answered using a mobile phone. Also, recent improvements in display quality and zooming have found applications in health sciences. Today, several software packages allow manipulation and internet transfer of radiological images using different protocols [18].

Touch screen technology has gained wide acceptance and is used in mobile phones, IPods, music players, and other devices [19]. In addition, some mobile phone companies including Apple and Samsung have developed internet-capable mobile phones that use this technology. This does not make the technology inaccessible to the blind. For example, "Voice Over" and "Siri" from Apple are voice services that read screen content and execute some spoken orders [20].

## 2.5. Wizards and privacy systems

In most current social network sites, users can customize privacy settings and policies. They may restrict access to photographs, videos, or other personal data [21]. Several techniques are available to simplify systems used to select privacy settings. Some ideas have come from work with filtering systems. In [22] authors defined collaborative filtering as a process of filtering based on opinion. For example, the latest movies may be sorted by evaluation and each movie may be ranked (1–5 stars). Any user can input a ranking and the overall opinion will be calculated and displayed. Recently, several websites have used this approach to optimize their sites by presenting the most important news based upon user evaluations. Such sites can suggest other information that users might like to view. Three types of collaborative filtering systems are recognized:

- Recommended items: If users like an item, the system will present other similar items for evaluation and possible purchase. For example, Amazon suggests items similar to recently purchased items.
- Predicted items: Specific items of interest are identified by calculating predicted ratings based on user input. This system is more popular than the recommendation system.
- Constrained recommendations: The items shown come with constraints, for example a list of all movies suitable for children.

In evaluating a "recommender" system questions such as the following are important: Does the system work? Are items that meet average ratings indeed recommended [22]? As an example of the use of a recommender system in the privacy context, Abdesslem and others [23] explored mobile location sharing. The system used the recommender technique to control personal information ("Where am I now"?) based on the collected behavioural data of friends. The system allows the user to disable the location service during specified times (for example, at night) and to hide location data from parents.

In [24] authors suggested development of a wizard allowing users to decline to share their locations with others. For example, in a university campus, a user can allow some other users, but not all, to locate him or her, then or later. Further, Shehab, Mohamed and Touati [25] developed a privacy system enabling a user to choose different privacy settings for each friend. A user can allow or deny his/her friend access to some personal data. The system was about 90% effective, but the list of friends studied was small and it is worth investigating whether this system would work if a user has a large number of friends?

## 2.6. Design of privacy systems

In social networking, the concept of privacy is divided into several sub-concepts all of which focus on attack and defence [26]. As shown from figure 1, it is important to assist users to set privacy policies. This allows the user to know who is seeing his/her personal data and to control data visibility. Also, the user should be informed of his/her privacy rights and the consequences of leaking personal information.

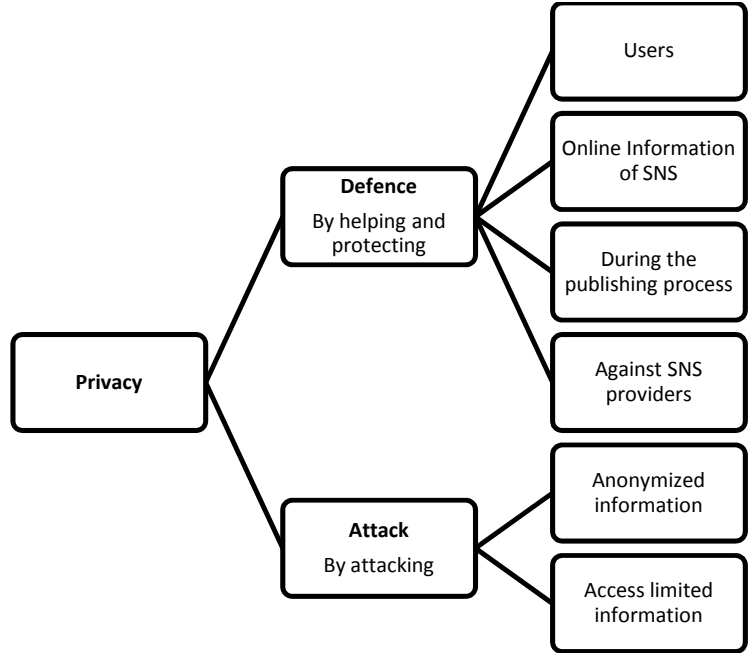

Figure 1. Privacy components and ramifications

Several privacy systems have been designed to protect personal information. For example, in [27] authors developed a framework enhancing privacy issues in identity management by introducing a middle-ware privacy level to give users more control of personal information. The new level could be set by users. The simplicity of some network tools such as addition of friends, uploading of photos, or commenting, has increased the chance of (accidentally) adding anonymous friends [28]. These authors found that use of privacy settings could reduce such risks. Further, Kolter and Pernul [29] emphasized that design simplicity, especially of the interface and tools of a privacy program, allowed users to optimally protect personal information. These authors used red, yellow, and green to indicate high-, medium-, or low-level privacy. This article will focus on assisting users to determine their privacy setting and will develop system useful in this regard.

## 3. OBJECTIVES

In this study, we develop a privacy system in the form of a wizard which allows users of internet mobile devices and PCs to control personal information privacy settings easily and quickly. We also aim to increase user awareness of privacy settings by allowing them to clarify exactly what information they want to share. In addition, we offer two different methods (default and custom) of controlling privacy settings. The system was tested by 439 volunteers, and the accuracy of the system was calculated. The wizard can be used to develop a new personal privacy system. In addition, we assessed user levels of concern about misuse of personal information by asking whether they currently hide such information.

However, this study differs from most prior works on personal information privacy in four key respects. First, the wizard affords both usability and flexibility. The wizard was built after conducting a pre-survey. Second, the system was trialled by 439 participants. Third, the accuracy of the system was assessed and it is clear that the system is valuable and can be developed further. Lastly, the simplicity of the design makes it possible to choose privacy settings on various types of mobile internet devices including those that use touch screen technology. The data will be used to develop a second-generation privacy system that is more refined.

## 4. METHODOLOGY

### 4.1. The smart wizard

The wizard's window allows a user to easily set privacy settings and save them to the server. The user is asked whether he/she wishes to show or hide personal details. The wizard simulates the results. The questions were designed based on pre-survey data [31]. When answers are given, the system sets a value of either 0 (meaning the information will be hidden) or 1 (the information will be visible). Figure 2 shows the general structure of the system.

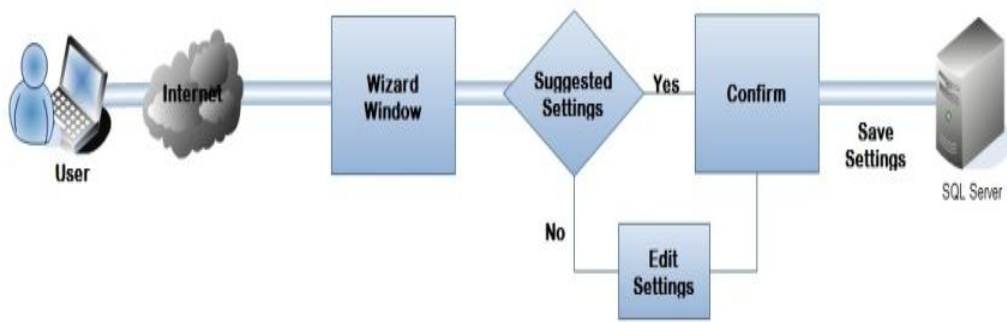

Figure 2. General structure of the wizard.

As can be seen from Figure 2, the system has several stages:
1.      The user logs in to the wizard webpage using the internet connection of his/her mobile phone or computer.
1.The wizard asks the user questions to which the user answers yes or no. The questions differ with respect to user choice.
2.  When all questions are answered, the system runs a privacy scan, based on the user's answers and the system then suggests privacy settings.
3.  The user can either hide or show any privacy item, or can confirm the proposed privacy profile in general.
4.  The user confirms and saves the chosen privacy settings

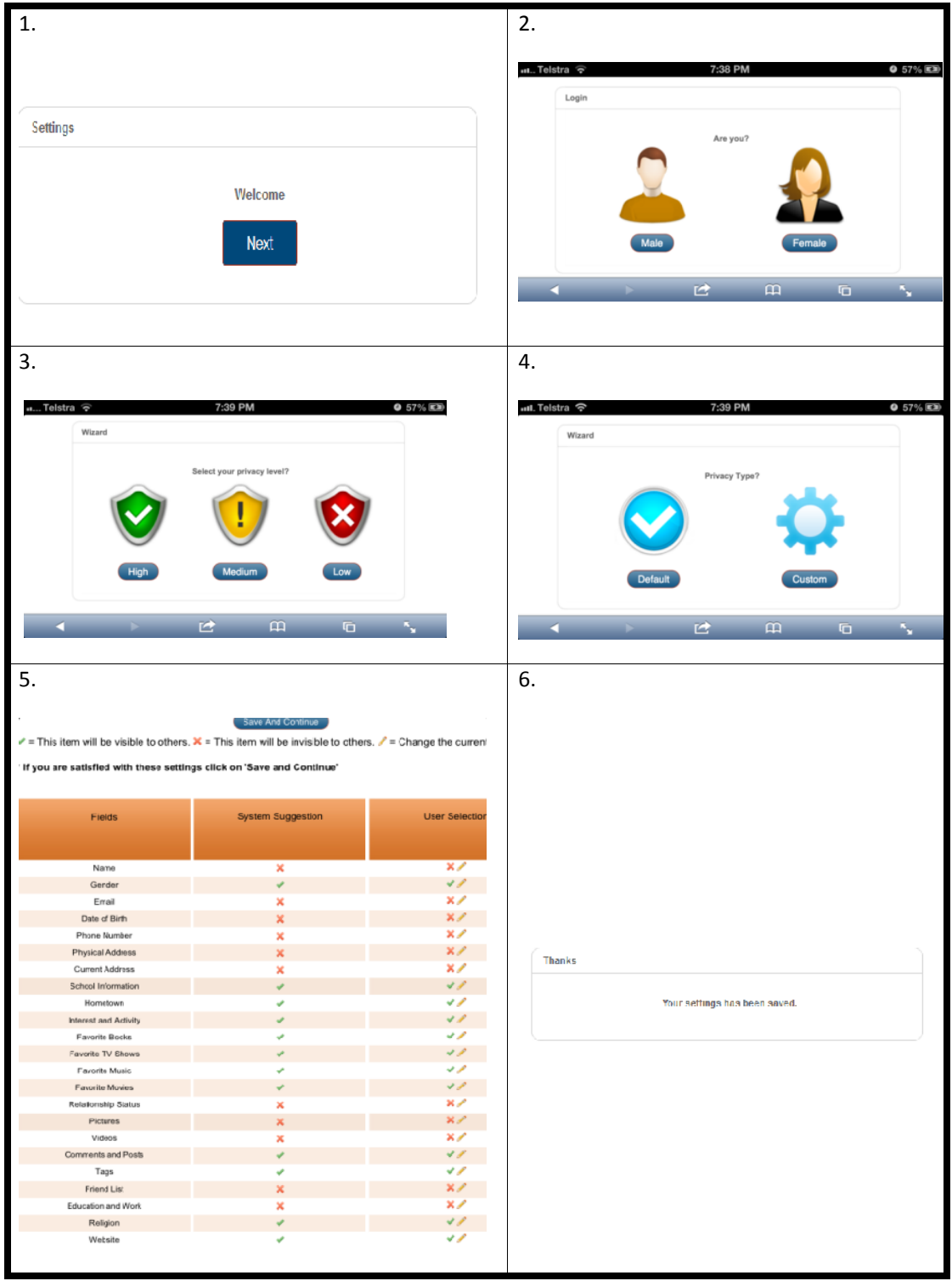

Figure 3. The female, medium, and default privacy profiles created by the wizard.

As shown in Figure 3, the wizard offers but a few options and page size is suitable for viewing on the screens of mobile devices. Also, touch screen technology is supported. The privacy levels (high, medium, and low) were chosen with reference to the results of the pre-survey [31]. A total of 185 volunteers of both sexes evaluated 23 items of personal information and ranked them from 1 to 5 (where 1 indicated low and 5 high privacy priority). We calculated mean values for all items, for all participants and by gender, and created three levels of privacy. Each level has two options: default or custom. The default settings reflect the pre-survey results, but, to provide freedom of choice, several questions ask the user if he/she wants to hide or show certain personal details. This allows the user to create custom settings. Figure 4 shows some of the questions posed when creating custom settings.

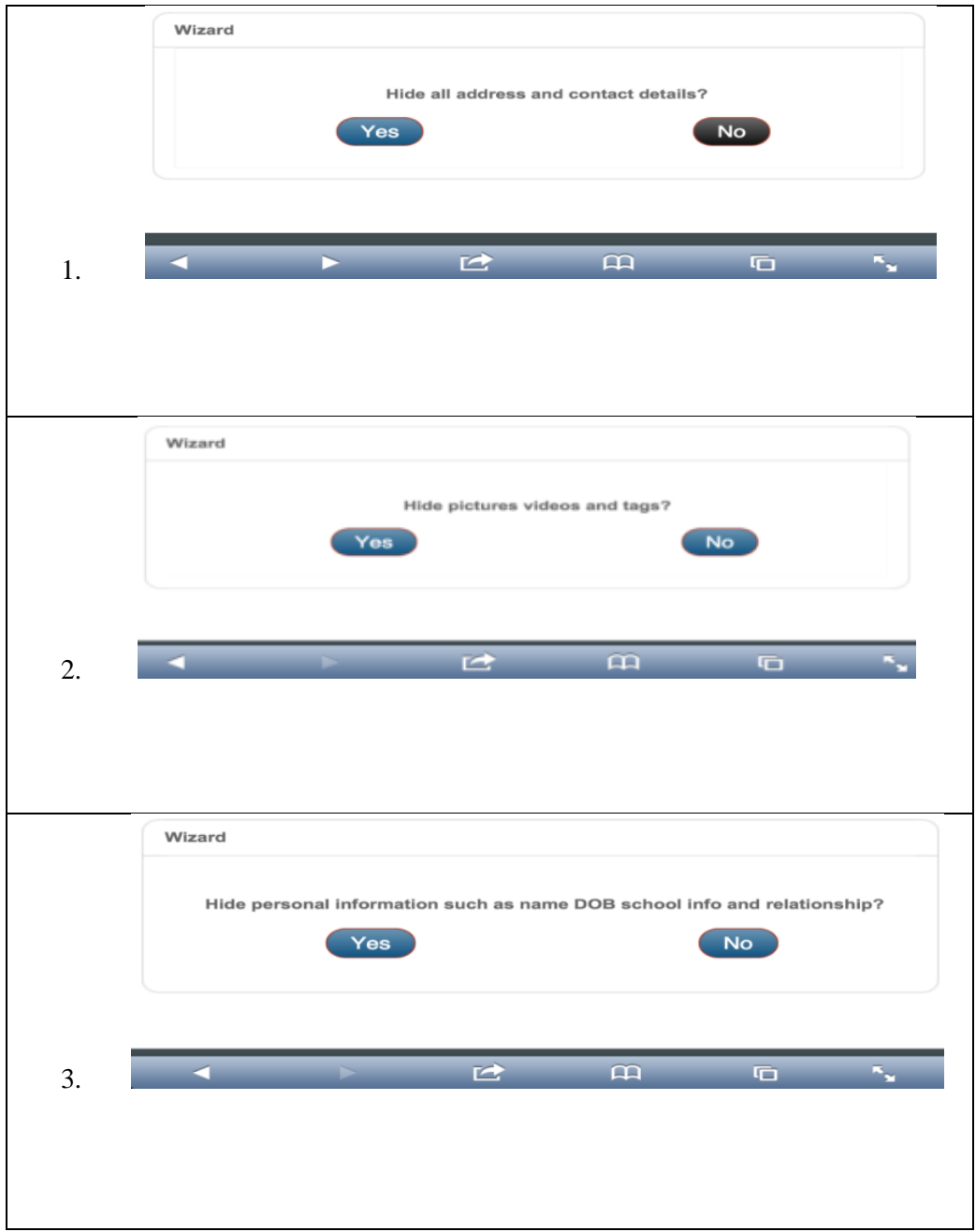

Figure 4. Custom privacy setting using the wizard.

## 4.2. Implementation of the wizard

### 4.2.1. Survey

The survey questionnaire was prepared in English and Arabic and used the ASP.net language and the SQL server 2008 structure. The target group was students and academic staff from various universities. An email invitation was sent to each member on staff and student email lists after approval from the University of New England; a covering letter describing the purpose of the survey, and the wizard, was included. The survey consisted of two different parts, of which the first explored system implementation and the second was a series of multiple choice questions exploring the concerns of users with respect to misuse of their personal data and whether they currently hid these data or not.

### 4.2.2. Reliability

Calculation of system accuracy is important because this allows the success rate of the wizard to be measured and determines if modifications are needed. Accuracy was calculated by estimating the number of personal information items that had been changed, using the edit window, and comparing these values to those suggested by the system.

To calculate accuracy for user e, the following formula was used:
*Accuracy $_e$ = (X ×100) / Y,*
where:
X = the number of personal information items that were not changed by the user, and
Y = the total number of all personal information items.
After calculating the accuracy for each user, the average accuracy for all participants was also calculated.

Mean Accuracy = $\dfrac{\sum_{e=0}^{|F|} \textit{Accuracy }_{F(e)}}{|F|}$

where F is the total number of users,

## 5. SURVEY RESULTS

## 5.1. The accuracy of the wizard

A total of 439 volunteers (352 males and 87 females) implemented the wizard and both languages were used (86 volunteers used English and 353 used Arabic). System accuracy was calculated as described above. The results were as follows

- Mean accuracy for both genders: 98.4%
- Mean accuracy for males: 98.13%
- Mean accuracy for females: 99.05%

## 5.2. Personal privacy

Table 1 shows the important features of personal information items. Some items were of high sensitivity. For example, personal videos, photographs, and tags were considered rather private and participants wished to hide them. In contrast, some items were not privacy-sensitive; these included favourite books, TV shows, and music. In this section, these issues will be explored further and between-gender differences discussed

Table 1. The relative importance of privacy.

| Item | The percentage of respondents willing to show the item | | |
|---|---|---|---|
| | Total | Male | Female |
| Name | 78.80% | 95.50% | 11.50% |
| Gender | 92.90% | 98.60% | 70.90% |
| Email | 4.10% | 4.50% | 2.30% |
| Date of birth | 77.90% | 95.70% | 5.70% |
| Phone number | 2.30% | 2.60% | 1.10% |
| Physical address | 3.20% | 3.70% | 1.10% |
| Current address | 3.20% | 3.70% | 1.10% |
| School information | 81.80% | 96.30% | 23% |
| Hometown | 90.90% | 97.40% | 64.30% |
| Interests and activities | 92.70% | 99.40% | 65.50% |
| Favourite book | 93.60% | 99.40% | 70.10% |
| Favourite TV show | 93.60% | 99.40% | 70.10% |
| Favourite music | 93.60% | 99.40% | 70.10% |
| Favourite movie | 93.60% | 99.40% | 70.10% |
| Relationship status | 77.40% | 95.20% | 5.70% |
| Pictures | 0% | 0% | 0% |
| Videos | 0% | 0% | 0% |
| Comment and posts | 99.10% | 100% | 95.40% |
| Tags | 0.70% | 0% | 3.45% |
| Friend list | 2.50% | 2.80% | 1.10% |
| Education and work | 76.80% | 94.60% | 4.60% |
| Religion | 92.70% | 99.40% | 65.50% |
| Website | 82.70% | 95.60% | 26.40% |

With respect to males, Figure 5 shows that 15 of 23 personal information items were of low sensitivity; they could be shared and misuse was not feared. However, 8 specific items, including the email address, the phone number, and the physical address, should often not be shared. This was also true of photographs, videos, and tags such as pictorial identification which might be misused.

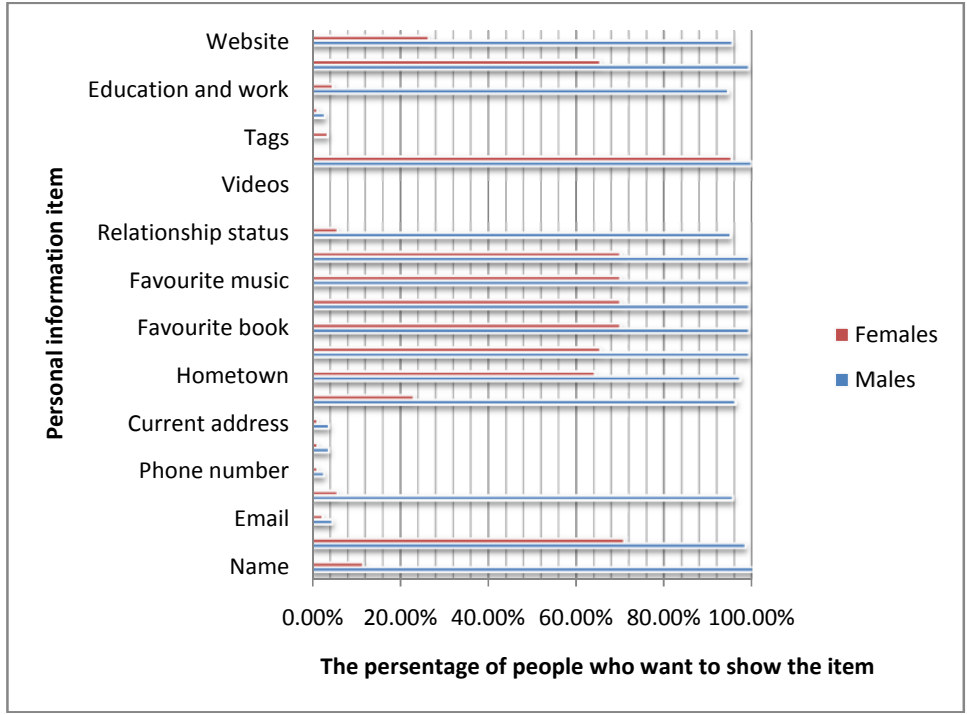

Figure 5. Privacy sensitivities of males and females.

Turning to females, Figure 5 shows that women were more careful about privacy than men. The willingness to share was lower than exhibited by males; fewer items were to be visible. Moreover, over 90% of respondents identified 11 items that should be hidden; these included their email address, phone number, and the physical address. Also, unlike males, name, relationship status, and education/work profiles were considered sensitive by females. In contrast, 60–71% of respondents (thus fewer than the proportions of males) identified items that could be shared. About 95% of females were willing to share comments and posts which, by their nature, are intended to be communicated. In general, females were more concerned with privacy than were males, especially in terms of items allowing personal identification.

## 5.3. Why hide personal information?

We assessed user awareness of misuse of personal information and whether it explained why users chose to hide some items. A total of 205 participants completed this part of the survey (131 males and 74 females). Table 2 reveals a close relationship between user concerns about misuse and the items hidden. For example, about 61% of participants were concerned about misuse of email addresses and about 51% hide these addresses. Moreover, favourite movies or music were of little concern because these items do not allow of personal identification or contact.

Table 2. Personal information items that concerned both genders and that were hidden.

| Item | Both genders | | Male | | Female | |
|---|---|---|---|---|---|---|
| | Concerned | Hide | Concerned | Hide | Concerned | Hide |
| Name | 51.22% | 41.46% | 44.27% | 35.88% | 63.51% | 51.35% |
| Gender | 20.49% | 19.02% | 16.8% | 14.5% | 27.02% | 27.02% |
| Email | 61.46% | 52.2% | 49.62% | 45.8% | 82.43% | 63.51% |
| Date of birth | 37.07% | 36.1% | 27.48% | 29.01% | 54.05% | 48.65% |
| Phone number | 74.63% | 72.68% | 65.65% | 65.65% | 90.54% | 85.14% |
| Physical address | 63.41% | 62.44% | 54.2% | 54.96% | 79.73% | 75.66% |
| Current address | 63.9% | 60.49% | 56.49% | 55.73% | 77.03% | 68.92% |
| School information | 22.44% | 24.4% | 17.5% | 22.14% | 31.08% | 28.38% |
| Hometown | 24.4% | 22.93% | 19.85% | 19.08% | 32.43% | 29.73% |
| Interests and activities | 17.7% | 15.61% | 12.98% | 10.69% | 24.32% | 24.32% |
| Favourite book | 14.14% | 14.63% | 9.92% | 10.69% | 21.62% | 21.62% |
| Favourite TV show | 12.68% | 13.66% | 9.16% | 9.92% | 18.92% | 20.27% |
| Favourite music | 14.15% | 13.17% | 10.69% | 9.92% | 20.27% | 18.92% |
| Favourite movie | 15.12% | 15.61% | 12.21% | 12.21% | 20.27% | 21.62% |
| Relationship status | 32.68% | 33.17% | 25.95% | 27.48% | 44.59% | 43.24% |
| Pictures | 61.95% | 55.61% | 52.67% | 52.67% | 78.38% | 60.81% |
| Videos | 62.93% | 56.59% | 54.96% | 53.44% | 77.03% | 62.16% |
| Comments and posts | 32.2% | 28.78% | 23.66% | 22.9% | 47.3% | 39.19% |
| Tags | 33.66% | 31.22% | 29.24% | 29.01% | 43.24% | 35.14% |
| Friends list | 53.66% | 44.88% | 45.8% | 41.22% | 67.57% | 51.35% |
| Education and work details | 26.34% | 25.37% | 19.08% | 19.85% | 39.19% | 35.14% |
| Religion | 20% | 19.51% | 16.03% | 16.03% | 27.03% | 25.68% |
| Website | 40.49% | 35.61% | 32.82% | 32.06% | 54.05% | 41.89% |

Figure 6 shows personal items of concern that were hidden by both males and females. The variables converged substantially because most users who are concerned about privacy and misuse of personal information hide such information. Some items were of high sensitivity (phone number, address, and personal photographs) and some of low sensitivity (hometown, favourite music, and others).

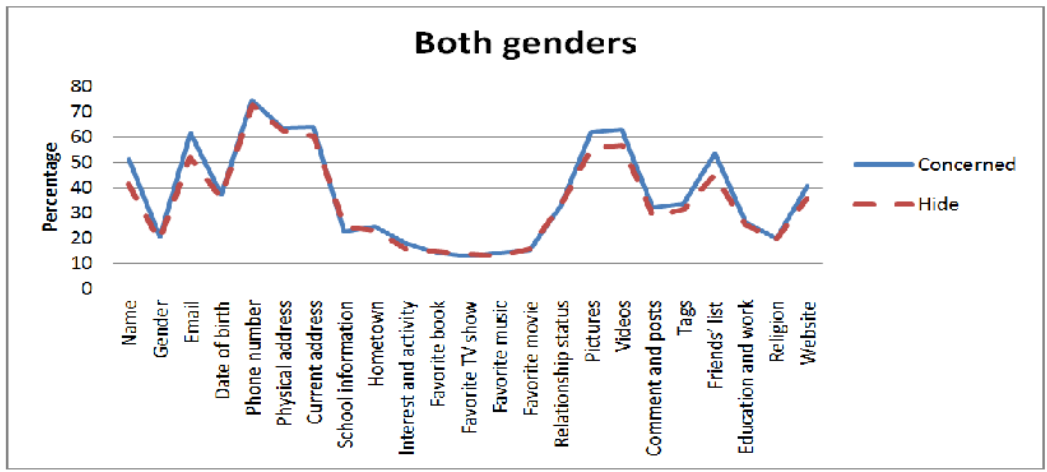

Figure 6. Combined male and female views of personal information that should or must be hidden.

Figure 7 presents data for males. In general, user concern was reflected in the hiding of items. Males were sensitive to personal photographs and videos, phone numbers, physical addresses, and current addresses. However, general topics such as favourite music and movies were not regarded as sensitive.

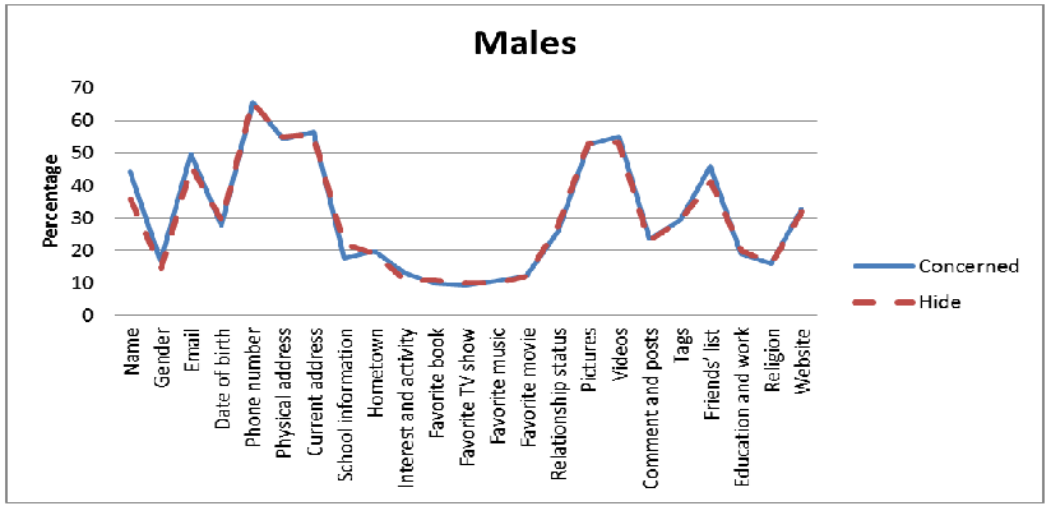

Figure 7. Male views of personal information that should or must be hidden.

Figure 8 shows the data for females. In comparison with Figure 8, some differences are evident. In general, females are more careful about privacy of personal information than are males. For example, only about 65% of males hid mobile phone information; the figure for females was about 90%. Moreover, the degree of sensitivity toward some items varied by gender. For example, name was of medium sensitivity for males (about 50%) but of higher sensitivity (about 63%) for females. However, although such differences were evident, both genders agreed that some items of personal information should be hidden.

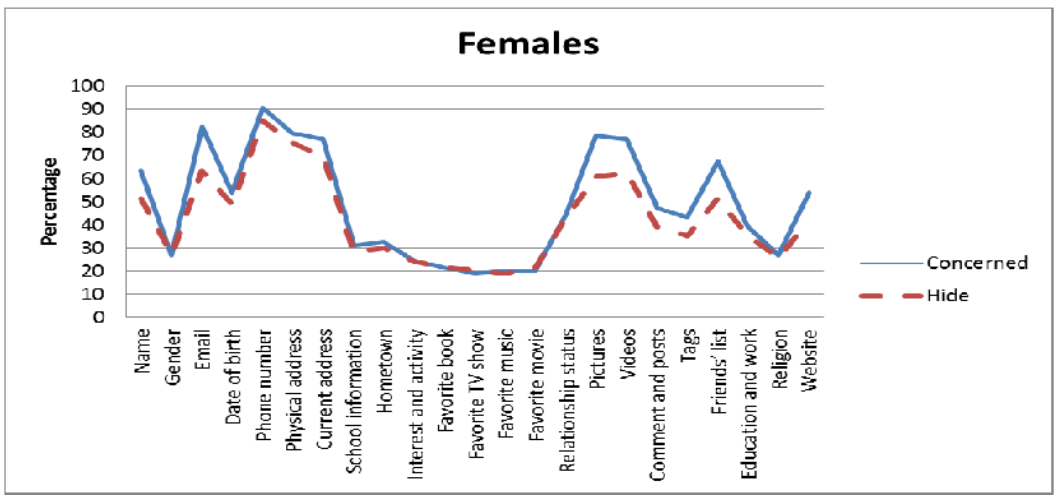

Figure 8. Female views of personal information that should or must be hidden.

## 6. LIMITATIONS OF THE RESEARCH

The study suffers from the quality of the composition of the sample. In Part 1, most survey participants (80%) were male but, in Part 2, only 64% were male.

The accuracy calculations are likely to be affected by respondent fatigue bias [32]. This occurs if a participant gets fatigued when asked to complete a large amount of tasks, which in our case was selecting their privacy choices for 23 items.

## 7. FUTURE WORK

The survey results show that the wizard worked very well; developers can build similar systems to help users modify privacy settings. Next, we will work to design a privacy system that takes a holistic view of privacy.

## 8. CONCLUSION

This paper has explored various aspects of the privacy of personal information and the ease with which privacy settings can be controlled. The widespread use of online social networks and mobile internet browsing encourage developers to design attractive applications. Increasing the numbers of advertisements is the main goal of developers, but protection of user privacy is a big challenge. In this article, we describe the development of a privacy wizard that helps users control privacy settings. The system was trialled by 439 users and the satisfaction rating was about 98%. Compared to males, females recorded more items of personal information to be held private. Further, the wizard effectively supports various types of mobile devices and makes users aware of the personal information they are sharing with others.

## ACKNOWLEDGEMENTS

In the beginning we would like to thank employees and students who participated with us. Also, we would like to thank all shareholder members of the university of Dammam and University of New England for facilitating this study. Finally, we would like to offer a special thank for Mr. Eid and Mr. Raddad for their assistance to collect the data.

## Authors

**Nahier Aldhafferi**

Nahier is a PhD student at University of New England in Australia. He works as a Lecturer at school of computer science in University of Dammam in Saudi Arabia. He holds a master degree in Internet Technology from University of Wollongong, Australia. Also, he is a lecturer at CISCO academy.

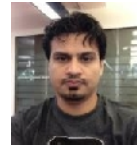

**Dr. Charles Watson**

Dr. Charles Watson is a UNE Council Member. He is a Convener of the Information Research Group and Faculty A&S Representative on e-Research Working Group. Also he is a Computer Science Representative on School S&T Research Committee. He lectures Computer Science at the University of New England in Australia, including units in networks, security, e-commerce and forensic computing. He was previously a Senior Research Scientist at DSTO and CSIRO and holds a PhD in Computer Science from the University of Sydney.

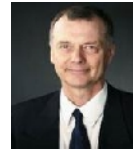

**Professor A. S. M. Sajeev**

Sajeev is the Chair in Computer Science/Information Technology at the University of New England. He holds a PhD in Computer Science from Monash University, Australia. His research interests are in software engineering: software metrics, testing, processes and project management. He also works in the area of mobile systems: interface design, language design, security and testing. He coordinates the IT Security theme within the University's targeted research area on security.

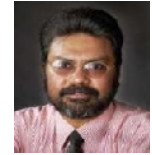